\documentclass[aps,pra,reprint]{revtex4-1}
\usepackage{amsmath,amsfonts,amssymb}
\usepackage{graphicx,float,calc}
\usepackage[breaklinks,colorlinks,urlcolor=blue,citecolor=blue,linkcolor=blue]{hyperref}

\def\bfr{{\bf r}}
\def\bfk{{\bf k}}

\begin{document}

\title{Cavity-induced Fulde-Ferrell-Larkin-Ovchinnikov superfluids of ultracold Fermi gases}
\author{Zhen Zheng}\thanks{zhenzhen.dr@outlook.com}
\affiliation{Department of Physics and HKU-UCAS Joint Institute for Theoretical and Computational Physics at Hong Kong, The University of Hong Kong, Pokfulam Road, Hong Kong, China}

\author{Z. D. Wang}\thanks{zwang@hku.hk}
\affiliation{Department of Physics and HKU-UCAS Joint Institute for Theoretical and Computational Physics at Hong Kong, The University of Hong Kong, Pokfulam Road, Hong Kong, China}
\affiliation{Frontier Research Institute for Physics, South China Normal University, Guangzhou 510006, China}

\begin{abstract}

Motivated by recent experimental advances in ultracold atomic gases placed in cavities,
we study the influence of the atom-cavity coupling on the Fermi gases trapped in optical lattices.
By adiabatic elimination of the cavity photon field,
the atom-cavity coupling gives rise to effective long-range interactions.
It results in a variety of two-body scattering processes,
during which the atomic pairs can acquire an additional center-of-mass momentum.
This reveals the possibility of Fulde-Ferrell-Larkin-Ovchinnikov (FFLO) superfluids in which the atomic pairing momentum is nonzero.
By inspecting the phase diagram at the mean-field level,
we confirm that the FFLO superfluid phase coexists with the zero-momentum pairing,
and is the ground state that hosts the lowest energy.
Furthermore, the order parameter characterizing the nonzero-momentum pairing does not vanish as long as the cavity-induced interaction is present.

\end{abstract}

\maketitle

\section{Introduction}

Exploring quantum many-body physics is one of the main challenging tasks in the studies of ultracold atoms \cite{cold-atom-rmp-2008}.
In the past decades, controllable interactions in ultracold atomic gases have been achieved by exploiting well-developed techniques bashed on Feshbach resonances \cite{feshbach-rmp-2006,feshbach-rmp-2010}.
It facilitates ultracold atomic gases serving as a versatile platform for exploring many interesting physical phenomena \cite{quantum-simulation-rev-nphys,quantum-simulation-rev-sci},
such as the crossover from the Bardeen-Cooper-Schrieffer (BCS) superfluid state to Bose-Einstein condensate (BEC) of molecules \cite{fermi-gas-rmp-2008}.
With the aid of laser fields,
a series of artificial gauge fields can be engineered in real experiments \cite{raman-rmp-2011,raman-rev-2014}.
This makes it possible for exploring unconventional quantum many-body states with nontrivial features.
Among them, the Fulde-Ferrel-Larkin-Ovchinnikov (FFLO) superfluid state has attracted great interest in research on ultracold Fermi gases \cite{Kinnunen_2018}.

It has been known that, in the presence of attractive atomic interactions,
the Fermi gas enters a superfluid state, in which atoms with opposite spins form pairs in the same way as electrons do in superconductors of solids \cite{leggett2006quantum}.
The essential physics of the system is captured by the BCS theory.
Generally, atomic pairs in the fermionic superfluid state host a zero center-of-mass (COM) momentum,
due to the nesting of spin-balanced Fermi surfaces \cite{Fradkin-pdw-2012}.
In the 1960s, Fulde and Ferrell \cite{fflo-1st-ff}, and Larkin and Ovchinnikov \cite{fflo-1st-lo} proposed
that non-zero-momentum pairing may be formed to lower the system energy in the presence of spin species imbalance \cite{fflo-jpsj},
which is known as the FFLO state.
The order parameter that characterizes the non-zero-momentum pairing exhibits a spatial modulation.
This is the essential distinction from the ordinary zero-momentum pairing.
Though there is a lack of prominent evidences in condensed-matter physics,
various schemes have been proposed to synthesize and demonstrate 
this unconventional state in ultracold Fermi gases by using artificial fields
\cite{zz-soc-fflo,Wu-soc-fflo,Dong_2013-fflo,Hu_2013-fflo,Iskin-fflo-2015,zz-shaken-fflo,zz-driven-fflo,Ghosh-fflo-2016,Liu-pi-fflo-2016,Chan-fflo-2017,Poon-fflo-2017,Dutta-fflo-2017,dark-state-fflo,Zheng_2019-fflo}.

Recently, the capability of preparing ultracold atoms in cavities was also exploited in searching unconventional quantum states \cite{CQED-rmp-2013}.
It can generate effective atomic interactions mediated by cavity photons \cite{baumann2010dicke,landig2016quantum,guo-pra2012}
and exhibit nontrivial phenomena including the self-organization \cite{Piazza-self-order-2014,Cosme-self-order-2018,Benjamin-self-order-2018,Kroeze-self-order-2018,Guo-self-order-2019,ref-add-prb2018},
supersolidity \cite{leonard2017supersolid,Mivehvar-supersolid-2018,Zhang-supersolid-2013}, 
charge density wave (CDW) \cite{Sheikhan-cdw-2019,Colella_jpb2019},
pairing density wave (PDW) \cite{Schlawin-pdw-2019},
higher-wave pairing \cite{schlawin2019cavity}, 
and magnetic order \cite{Fan-magnetic-2018,Colella_njp2019}.
Based on these pieces of previous work,
the spatially modulated structure inspires us to inspect the influence of the cavity-induced interaction on the fermionic superfluid phase,
which may raise the possibility of the emergent FFLO state.

The paper is organized as follows.
In Sec.\ref{sec-model}, we consider the model Hamiltonian describing fermionic atoms coupled to a pumped optical cavity,
and present the form of the cavity-induced long-range interaction.
In Sec.\ref{sec-pairing}, we find that the long-range interaction will involve various two-body scattering processes of atoms.
It leads to a distinguished pairing mechanism between atoms of opposite spins,
and reveals the existence of non-zero-momentum pairing.
The numeric results are shown in Sec.\ref{sec-numerics}.
In Sec.\ref{sec-discussion}, we discuss the extension of the model Hamiltonian in the one-dimensional (1D) lattice.
In Sec.\ref{sec-conclusion}, we summarize this work.

\section{Model Hamiltonian} \label{sec-model}

\begin{figure}[t]
\centering
\includegraphics[width=0.49\textwidth]{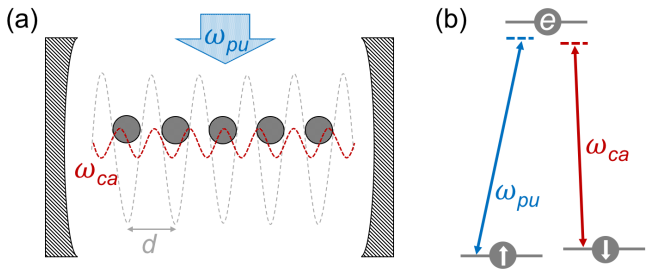}
\caption{Experimental setups:
(a) Atoms of the ultracold Fermi gas are coupled to the cavity field ($\omega_{\rm ca}$)
and the pumping laser field ($\omega_{\rm pu}$) perpendicular to the cavity.
The grey dashed line illustrates the trap potential of the optical lattice.
(b) Sketch of the atomic transition driven by the cavity and pump laser fields.}
\label{fig-setup}
\end{figure}

We start from an ultracold Fermi gas trapped in a three-dimensional (3D) optical lattice.
Two hyperfine states of the atoms can be chosen
as pseudospin $\uparrow$ and $\downarrow$.
The Fermi gas is prepared to be placed in a linear cavity oriented in the $x$ direction,
while a pumping laser is placed perpendicular to the cavity.
The experimental setup for our proposal is sketched in FIG.\ref{fig-setup}.
In this setup, the spin-$\uparrow\downarrow$ states can be coupled through an intermediate excited state $|e\rangle$.
Such a three-level Raman transition
is constructed by the single-mode standing-wave cavity-photon field (with frequency $\omega_{\rm ca}$)
assisted with the pumping laser field (with frequency $\omega_{\rm pu}$).
This setup comprises a driven-dissipative system due to the pumping laser and cavity-photon losses,
and the steady state of the system can be described by the Jaynes-Cummings model \cite{maschler2008ultracold}.
After adiabatically eliminating the excited state $|e\rangle$,
the steady state of the system is described by the following Hamiltonian composed of four terms:
\begin{equation} \label{eq-h-start}
H = H_{\rm a} + H_{\rm aa} + H_{\rm c} + H_{\rm ac} \,.
\end{equation}
The first and second terms $H_{\rm a} + H_{\rm aa}$ describe the atom subsystem.
In particular, $H_{\rm a}$ is the single-particle Hamiltonian of atoms,
\begin{equation} \label{eq-h-start-a}
H_{\rm a}=\int  d \bfr \sum_\sigma \psi_{\sigma}^\dag(\bfr) [ -\nabla^2/2m-\mu_\sigma + V_L(\bfr) ] \psi_{\sigma}(\bfr) \,,
\end{equation}
while $H_{\rm aa}$ describes the contact two-body interactions of atoms which is generally created by Feshbach resonances:
\begin{equation} \label{eq-h-start-aa}
H_{\rm aa}=-\int d\bfr d\bfr'\, g_{\rm fr}\delta(\bfr-\bfr') \psi_\uparrow^\dag(\bfr')\psi_\downarrow^\dag(\bfr)
\psi_\downarrow(\bfr)\psi_\uparrow(\bfr') \,.
\end{equation}
Here the trap potential of the optical lattice is $V_L(\bfr)=\sum_{l=x,y,z}V_L\sin^2(k_Lx)$ 
($k_L=\pi/d$ with $d$ standing for the lattice constant, and $V_L$ is the trap depth),
$m$ is atomic mass,
$\mu_\sigma$ is the chemical potential of spin-$\sigma$ atoms,
and $g_{\rm fr}$ is the bare strength of the contact interaction.

The third term $H_{\rm c}$ in Hamiltonian (\ref{eq-h-start}) describes the cavity subsystem,
\begin{equation} \label{eq-h-start-c}
H_{\rm c} = - \Delta_{\rm ca} a^\dag a \,,
\end{equation}
Here $\Delta_{\rm ca}=\omega_{\rm pu}-\omega_{\rm ca}$ is the cavity-pumping detuning.
We note that, in the driven-dissipative system,
one needs to account for the cavity losses phenomenologically via a decay rate $\kappa$.
In practice, this decay can be prepared to be vanishingly small ($\kappa\ll\Delta_{\rm ca}$) in the superradiant regime \cite{decay-diss}
and is neglected hereafter.

The last term $H_{\rm ac}$ in Hamiltonian (\ref{eq-h-start}) describes the interaction between the two subsystems:
\begin{equation} \label{eq-h-start-ac}
H_{\rm ac} = \int  d \bfr \,
g(x) [a\psi_\uparrow^\dag(\bfr)\psi_\downarrow(\bfr)+H.c.] \,.
\end{equation}
Here $g(x)=g\cos(k_cx)$ is the atom-cavity coupling mode,
and $H.c.$ stands for the Hermitian conjugation.
We suppose $g\ll V_L$;
thus, the atom-cavity coupling does not change the lattice configuration.
The detailed formulas for obtaining Eqs.(\ref{eq-h-start-c}) and (\ref{eq-h-start-ac}) are given in Appendix \ref{sec-app-model}.

In the atom-cavity system described by Hamiltonian (\ref{eq-h-start}),
the partition function is expressed as
\begin{equation}
\mathcal{Z} = {\rm Tr}[e^{-\beta H}] = \int \mathcal{D}\psi\mathcal{D}a \, e^{-\mathcal{S}[\psi,a]} \,,
\end{equation}
where $\mathcal{S}$ is the action of the system,
\begin{equation}
\mathcal{S}[\psi,a] = \int  d \tau  d \bfr \sum_\sigma\psi_\sigma \partial_\tau \psi_\sigma + H(\psi,a) \,.
\end{equation}
Here $\tau$ denotes the imaginary time and $\beta\equiv 1/T$ with $T$ as the temperature.
In order to obtain an effective Hamiltonian $H_{\rm eff}$ that solely characterizes the physics of the atom subsystem,
we adiabatically eliminate the cavity field $a$ by integrating them out in $\mathcal{Z}$.
Then the Hamiltonian $H(\psi,a)$ in $\mathcal{S}[\psi,a]$ is replaced by an effective one $H_{\rm eff}(\psi)$,
whose form is expressed as 
\begin{equation} \label{eq-h-eff-1}
H_{\rm eff} = H_{\rm a} + H_{\rm int},
\end{equation}
where the interacting part can be written as
$H_{\rm int} = H_{\rm aa}+\widetilde{H}_{\rm ac}$
with
\begin{equation} \label{eq-h-int-ac}
\widetilde{H}_{\rm ac}=
\int  d\bfr d\bfr'\,\frac{g(x)g(x')}{\Delta_{\rm ca}}\psi_\uparrow^\dag(\bfr)\psi_\downarrow(\bfr)
\psi_\downarrow^\dag(\bfr')\psi_\uparrow(\bfr') \,.
\end{equation}
Eq.(\ref{eq-h-int-ac})
reveals that the adiabatic elimination of the cavity field gives rise to an effective long-range interaction, 
besides the contact one ($H_{\rm aa}$) created by Feshbach resonances.
Moreover, the form of the effective interaction is modulated in real space.
By making the Fourier transformation, the form of $H_{\rm int}$ in momentum space is expressed as
\begin{equation} \label{eq-h-int-k-total}
H_{\rm int} = H_{\rm int}^{(0)}+H_{\rm int}^{(1)}+H_{\rm int}^{(2)}+H_{\rm int}^{(3)} \,,
\end{equation}
where
\begin{equation} \label{eq-h-int-k-each}
\left\{\begin{split}
& H_{\rm int}^{(0)} = U_{\rm ac}\sum_{\bfk} 2\psi_{\bfk\downarrow}^\dag\psi_{\bfk\downarrow}+ 
(\psi_{\bfk+2\bfk_c\downarrow}^\dag\psi_{\bfk\downarrow} + H.c.)\\
&H_{\rm int}^{(1)}=-U_{\rm fr}\sum_{\bfk,\bfk',{\bf p}}
\psi_{\bfk'-{\bf p}\uparrow}^\dag \psi_{\bfk+{\bf p}\downarrow}^\dag\psi_{\bfk\downarrow}\psi_{\bfk'\uparrow}\\
&H_{\rm int}^{(2)}=-U_{\rm ac}\sum_{\bfk,\bfk'}\sum_{\zeta=\pm}
\psi_{\bfk-\zeta \bfk_c\uparrow}^\dag \psi_{\bfk'+\zeta \bfk_c\downarrow}^\dag\psi_{\bfk\downarrow}\psi_{\bfk'\uparrow}\\
&H_{\rm int}^{(3)}=-U_{\rm ac}\sum_{\bfk,\bfk'}\sum_{\zeta=\pm}
\psi_{\bfk+\zeta \bfk_c\uparrow}^\dag \psi_{\bfk'+\zeta \bfk_c\downarrow}^\dag\psi_{\bfk\downarrow}\psi_{\bfk'\uparrow}
\end{split}\right.\,.
\end{equation}
Here we have denoted $\bfk_c=(k_c,0,0)$.
The forms of $U_{\rm fr}$ and $U_{\rm ac}$ are given in Appendix \ref{sec-app-int}.

\section{Pairing Mechanism} \label{sec-pairing}

\begin{figure}[t]
\centering
\includegraphics[width=0.49\textwidth]{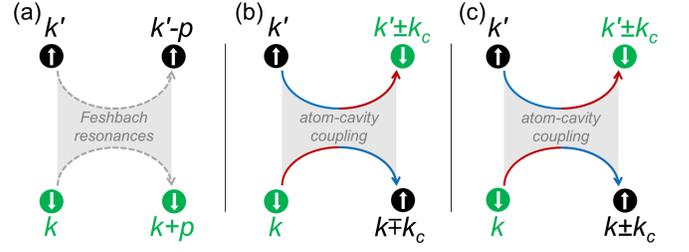}
\caption{Two-body scattering of atoms.
(a) Process governed by Feshbach resonances (described by $H_{\rm int}^{(1)}$ of Eq.(\ref{eq-h-int-k-each})).
(b)-(c) Processes governed by the atom-cavity coupling (described by $H_{\rm int}^{(2)}$ and $H_{\rm int}^{(3)}$, respectively).}
\label{fig-scatter}
\end{figure}

Now we inspect the interacting Hamiltonian (\ref{eq-h-int-k-each}).
The first term $H_{\rm int}^{(0)}$ can be merged into the term $H_{\rm a}$ of Eq.(\ref{eq-h-eff-1}),
since it describes a modification to the single-particle Hamiltonian of spin-$\downarrow$ atoms.
Furthermore, the first term of $H_{\rm int}^{(0)}$ can be recognized as a constant energy term,
which can be compensated by Stark shifts generated by optical fields (see Appendix \ref{sec-app-model})
and is therefore neglected hereafter.
In contrast to $H_{\rm int}^{(0)}$, the last three terms $H_{\rm int}^{(1\sim3)}$
describe two-body interactions between opposite spins.
During the two-body scattering governed by $H_{\rm int}^{(1)}$ and $H_{\rm int}^{(2)}$,
the COM momentum of atomic pairs is conserved, as shown in FIG.\ref{fig-scatter}(a)-(b).
However, during the scattering by $H_{\rm int}^{(3)}$, this process violates the momentum conservation.
Each atomic pair acquires an additional COM momentum of $\pm 2\bfk_c$ after the scattering.
This is because, during the Raman transition (see FIG.\ref{fig-scatter}(b)),
the atom-cavity coupling can introduce a net momentum transfer of $\pm 2\bfk_c$ to atomic pairs.

When $U_{\rm fr}, U_{\rm ac}>0$, the two-body interactions generated by $\mathcal{H}_{\rm int}^{(1\sim3)}$ are all attractive.
In this regime, the Fermi gas enters the fermionic superfluid state,
and we regard the atomic pairing as its order parameter.
For 3D ultracold Fermi gases,
the mean-field Bogoliubov-de Gennes (BdG) approach is a useful approximation that is widely applied in investigating the superfluid state.
Although the BdG theory is not expected to be quantitatively accurate,
it can still provide a comprehensive description and qualitatively capture the physics picture along the BCS-BEC crossover at zero temperature \cite{fermi-gas-rmp-2008,Dutta-mean-field-2016}.
According to the earlier discussions,
the atomic pairing hosts a COM momentum of zero or $\pm 2\bfk_c$ (see FIG.\ref{fig-scatter}).
Therefore, at the mean-field level, we introduce dimensionless order parameters $\Delta$ as follows:
\begin{equation} \label{eq-order}
\left\{\begin{split}
&U_{\rm fr}\langle \psi_{\bfk\downarrow} \psi_{\bfk'\uparrow} \rangle=
U_{\rm fr}\Delta_0\delta_{\bfk+\bfk',0} \\
&U_{\rm ac}\langle \psi_{\bfk\downarrow} \psi_{\bfk'\uparrow} \rangle=
U_{\rm ac}(\Delta_0\delta_{\bfk+\bfk',0} + \Delta_{2k_c}\delta_{\bfk+\bfk',2\bfk_c})
\end{split}\right.\,,
\end{equation}
in which the cavity-induced pairing yields a superposition of zero- and finite-momentum ones.
In Eq.(\ref{eq-order}), we have accounted for various pairing mechanisms.
$\Delta_0$ stands for the order parameter that characterizes ordinary zero-momentum pairing.
By contrast, a nonzero $\Delta_{2k_c}$ yields the FFLO superfluid phase associated with finite-momentum pairing.

In the presence of the periodic lattice potential,
we use the tight-binding approximation to study the system.
For simplicity, we can prepare the optical lattice and cavity mode to make $2k_c$ equal to $k_L$.
In this way, the BdG Hamiltonian can be cast into a matrix form with finite dimensions.
We choose the base $\widetilde{\Psi}_\bfk = (\Psi_\bfk,\Psi_{-\bfk}^\dag)^T$ with
$\Psi_\bfk= (\psi_{\bfk+\bfk_L/2\uparrow},\psi_{\bfk+\bfk_L/2\downarrow},\psi_{\bfk-\bfk_L/2\uparrow},\psi_{\bfk-\bfk_L/2\downarrow})$ and $\bfk_L=(k_L,0,0)$.
The BdG Hamiltonian is thus written as
\begin{equation} \label{eq-h-bdg}
H_{\rm BdG}(\bfk) =
\begin{pmatrix}
H_0(\bfk) & \hat{D} \\
\hat{D}^\dag & -H_0(-\bfk)
\end{pmatrix} \,.
\end{equation}
Here the single-particle part reads
\begin{equation}
H_0(\bfk)=
\begin{pmatrix}
\xi_{\bfk+\bfk_L/2\uparrow} & 0 & 0 & 0 \\
0 & \xi_{\bfk+\bfk_L/2\downarrow} & 0 & U_{\rm ac}/2 \\
0 & 0 & \xi_{\bfk-\bfk_L/2\uparrow} &0 \\
0 & U_{\rm ac}/2 & 0 & \xi_{\bfk-\bfk_L/2\downarrow}
\end{pmatrix}
\end{equation}
and the pairing part is given by
\begin{align}
\hat{D}&= 2U_{\rm ac}(\Delta_0 + \Delta_{k_L}) (\sigma_0+\sigma_x)\otimes(i\sigma_y) \notag\\
&+U_{\rm fr}\Delta_{0}\sigma_x\otimes(i\sigma_y) \,.
\end{align}
In $H_0(\bfk)$, $\xi_{\bfk\sigma}=-2t\sum_{l=x,y,z}\cos(k_ld)-\mu_\sigma$
represents the band dispersion of the tight-binding model.
$t$ is the hopping magnitude between adjacent lattices, 
which we choose as the energy unit hereafter.
In the matrix $\hat{D}$, $\sigma_{0,x,y,z}$ are Pauli matrices.

The thermodynamic potential of system can be calculated by \cite{Hu-fflo-2006}
\begin{equation}
\Omega =\mathcal{S}/\beta= \mathcal{E}_{0} + \sum_{\bfk,\sigma}\frac{\xi_{\bfk\sigma}}{2}
-\frac{1}{4\beta}\sum_{\bfk,n} \ln(1+e^{-\beta E_\bfk^n}) \,.
\end{equation}
Here $\mathcal{E}_{0} = 2U_{\rm ac}|\Delta_0+\Delta_{k_L}|^2+U_{\rm fr}|\Delta_0|^2$,
and $E_\bfk^n$ ($n=1\sim8$) are eigenvalues of the BdG Hamiltonian (\ref{eq-h-bdg}).
For the ground state of the system,
the order parameters $\Delta_0$ and $\Delta_{k_L}$ are determined by minimizing the thermodynamic potential $\Omega$.
It can be achieved by self-consistently solving the following equations:
\begin{equation}
\frac{\partial}{\partial\Delta_0}\Omega=\frac{\partial}{\partial\Delta_{k_L}}\Omega=0 \,.
\end{equation}

\section{Numeric results} \label{sec-numerics}

\begin{figure}[t]
\centering
\includegraphics[width=0.49\textwidth]{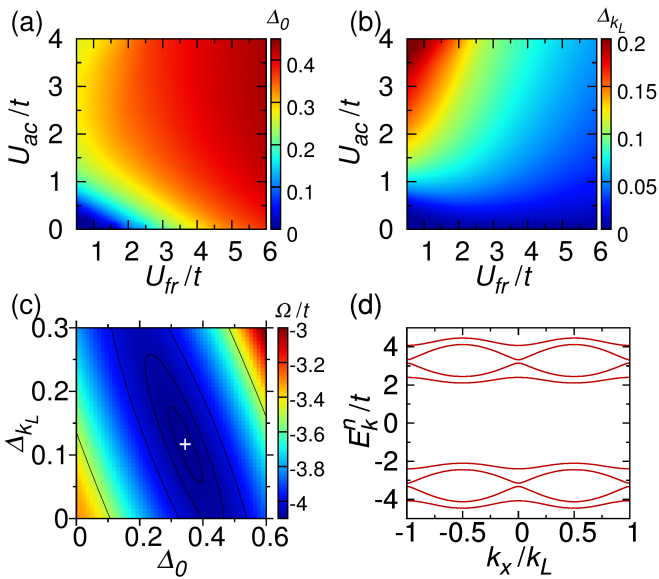}
\caption{Order parameters as functions of $U_{\rm fr}$ and $U_{\rm ac}$.
The color describes the amplitude of (a) $\Delta_0$ and (b) $\Delta_{k_L}$.
We set $\mu_\uparrow=\mu_\downarrow=1.0t$.
(c) The contour plot of the thermodynamic potential $\Omega$ in the $\Delta_0$-$\Delta_{k_L}$ plane 
for $(U_{\rm fr},U_{\rm ac})=(2.0,3.0)t$.
The white cross symbol corresponds to the global minimal energy.
(d) Quasiparticle band structure for (c) at $k_y=k_z=0$.
$E_\bfk^n$ ($n=1\sim8$) are eigenvalues of Hamiltonian (\ref{eq-h-bdg}).}
\label{fig-phase}
\end{figure}

In ultracold atoms, the strength of the contact interaction and the effective long-range one (i.e., $U_{\rm fr}$ and $U_{\rm ac}$)
can be controllable via magnetic or optical fields.
In FIG.\ref{fig-phase}(a) and (b), 
we respectively display the order parameters $\Delta_0$ and $\Delta_{k_L}$ in the $U_{\rm fr}$-$U_{\rm ac}$ plane
at zero temperature.
We can see that $\Delta_0$ increases monotonically 
with respect to both $U_{\rm fr}$ and $U_{\rm ac}$.
When $U_{\rm ac}=0$, $\Delta_{k_L}$ vanishes.
This is consistent with the actual physics in the BCS-BEC crossover,
because in the absence of the cavity field, 
the system reduces to the conventional BCS superfluid phase, which has been widely studied \cite{fermi-gas-rmp-2008}.

With the increase of $U_{\rm ac}$, $\Delta_{k_L}$ increases monotonically from zero.
This is because in Hamiltonian (\ref{eq-h-int-k-each}), 
$H_{\rm int}^{(3)}$ drives a transition from the zero-momentum pairing to the nonzero-momentum one,
by enforcing a transfer of $\pm 2\bfk_c$ to the atomic pairs.
It indicates that the effective long-range interaction governed by $H_{\rm int}^{(3)}$ excites and stabilizes atomic pairs with COM momentum of $\pm2\bfk_c$.
In FIG.\ref{fig-phase}(b), we can see that
even though $U_{\rm ac}$ is extremely weak,
the finite-momentum pairing, estimated by $\Delta_{k_L}$,
can still be excited from the zero-momentum one via the weak cavity-induced interaction.
In other words, the system enters the FFLO superfluid phase as long as $U_{\rm ac}\neq0$.
It is the ground state that hosts the lowest energy in the global order-parameter plane,
as shown in FIG.\ref{fig-phase}(c).
The hole bands are fully gapped from the particle bands, as shown in FIG.\ref{fig-phase}(d), and thus the superfluid state can be robust against small thermal fluctuations.

We remark that, since $\Delta_0$ and $\Delta_{k_l}$ are simultaneously nonzero,
the magnitude of the total order parameter will be staggered as $\Delta_0+(-1)^{x_j}\Delta_{k_l}$ 
($x_j$ is the $x$ coordinate of the $j$-th site).
It performs a stripe structure along the $x$ direction.

\section{Discussion} \label{sec-discussion}

In this paper,
we have focused on the case that the momentum transfer $2k_c$ during the Raman transition 
matches the reciprocal lattice vector $k_L$.
The two-body scattering illustrated by FIG.\ref{fig-scatter}(c) is thus indeed the well-known Umklapp process.
It brings in interesting phenomena for Fermi gases trapped in 1D optical lattices,
which simulate the Luttinger liquids
and can perform transitions from CDW to superconductor phases \cite{1D-rmp-2011} by changing the cavity-induced interaction.
We remark that in conventional solid systems,
the Umklapp process generally works when each site is half filled \cite{giamarchi2003quantum}.
By contrast in this paper, the momentum transfer $2k_c$ is tunable and can be set to match the Fermi wave vector $k_F$ of the 1D system.
Thus, the cavity-induced Umklapp process in this paper will be present even at an incommensurate filling of atoms.

When the Raman momentum transfer $2k_c$ equals to a generic form,
for instance, the fractal pattern $k_c = k_Lp/q$ ($p$ and $q$ are prime integers),
the pairing mechanism associated with nonzero COM momentum still works.
However, the energy band for each spin degree of freedom will be split into $q$ subbands.
Meanwhile, the atomic pairs will host various COM momenta of $\pm 2q'k_c$ ($q'=0,1,...,q$).

\section{Conclusion} \label{sec-conclusion}

In summary, we have investigated the influence of the atom-cavity coupling on the Fermi gases trapped in optical lattices.
It gives rise to an effective long-range interaction and involves a variety of scattering processes,
during which the atomic pairs will exhibit nonconserved COM momentum.
The non-zero-momentum pairing is thus naturally excited and coexists with the zero-momentum one,
resulting in the FFLO superfluid phase as the ground state of the lattice system.

\section{Acknowledgements}

This work was supported by the Key-Area Research and Development Program
of GuangDong Province (Grant No. 2019B030330001), the National Key Research
and Development Program of China (Grant No. 2016YFA0301800), the National Natural Science Foundation of China (No. 11704367),
the GRF (No.: HKU173057/17P) and CRF (No.: C6005-17G) of Hong Kong.

\appendix

\section{Modeling the Atom-Cavity Coupled Hamiltonian} \label{sec-app-model}

The atom-cavity coupled system is described by the Jaynes-Cummings Hamiltonian composed of three parts \cite{maschler2008ultracold},
\begin{equation} \label{eq-app-h-start}
\mathcal{H} = \mathcal{H}_{\rm c} + \mathcal{H}_{\rm a} + \mathcal{H}_{\rm Ra} \,.
\end{equation}
The first part $\mathcal{H}_{\rm c}$ describes the cavity photon energy,
\begin{equation}
\mathcal{H}_{\rm c}=\omega_{\rm ca} a^\dag a \,,
\end{equation}
where $\omega_{\rm ca}$ is the frequency of the cavity photon field.
We the pseudospin-$\uparrow$ level as the zero-energy state;
then the second part $\mathcal{H}_{\rm a}$ describing the single-particle Hamiltonian of the atoms is given by
\begin{equation}
\mathcal{H}_{\rm a} = \int  d \bfr \,[\Gamma_\downarrow \psi_{\downarrow}^\dag(\bfr) \psi_{\downarrow}(\bfr)
+ \Gamma_e\psi_{e}^\dag(\bfr) \psi_{e}(\bfr)] \,.
\end{equation}
Here we have neglected the kinetic motion of atoms at low temperature, 
since they are weak compared with the optical field strength.
$\Gamma_{\downarrow(e)}$ is the energy difference of the state $|\downarrow\rangle$ ($|e\rangle$) with respect to $|\uparrow\rangle$.
The last part $\mathcal{H}_{\rm Ra}$ describes the Raman transition driven by the pump laser and cavity photon fields (see FIG.\ref{fig-setup}(b)).
It is formulated as follows,
\begin{align}
\mathcal{H}_{\rm Ra} &= \int  d \bfr \, \Omega_{\rm pu}e^{-i\omega_{\rm pu}t} \psi_e^\dag(\bfr) \psi_\uparrow(\bfr) \notag\\
& +\Omega_{\rm ca}(x) a\psi_e^\dag(\bfr) \psi_\downarrow(\bfr) 
+ H.c. \,,
\end{align}
where $\omega_{\rm pu}$ is the frequency of the laser field,
$\Omega_{\rm ca}(x)=\Omega_{\rm ca}\cos(k_cx)$ is the cavity field mode,
and $\Omega_{\rm pu}$ and $\Omega_{\rm ca}$ are strengths of the pumping laser and cavity photon fields.

In order to obtain a time-independent effective Hamiltonian,
we make the following unitary rotation:
\begin{equation}
\mathcal{U} = \exp \big[ i\omega_{\rm pu}t\int\psi_e^\dag(\bfr)\psi_e(\bfr) d\bfr+i\omega_{\rm pu}t\, a^\dag a \big] \,.
\end{equation}
In the rotation frame, Hamiltonian (\ref{eq-app-h-start}) is rewritten as
\begin{align}
& \mathcal{H}=\mathcal{U} \mathcal{H} \mathcal{U}^\dag -  \mathcal{U} i\partial_t \mathcal{U}^\dag \\
&=-\Delta_{\rm ca} a^\dag a + \int  d \bfr \,\{ \Gamma_\downarrow \psi_{\downarrow}^\dag(\bfr) \psi_{\downarrow}(\bfr)
- \Delta_e\psi_{e}^\dag(\bfr) \psi_{e}(\bfr) \notag\\
& +[\Omega_{\rm pu} \psi_e^\dag(\bfr) \psi_\uparrow(\bfr)
+\Omega_{\rm ca}(x) a\psi_e^\dag(\bfr) \psi_\downarrow(\bfr) 
+ H.c.] \} \,, \label{eq-app-h-1}
\end{align}
where the detuning $\Delta_{\rm ca} = \omega_{\rm pu}-\omega_{\rm ca}$ and $\Delta_e=\omega_{\rm pu}-\Gamma_e$.
For the field $\psi_e$, its equation of motion is
\begin{equation}
i\partial_t \psi_e = [\psi_e, \mathcal{H}]= -\Delta_e\psi_e + \Omega_{\rm pu} \psi_\uparrow + \Omega_{\rm ca} a \psi_\downarrow \,.
\end{equation}
By adiabatically eliminating it, i.e., setting $\partial_t\psi_e=0$, we obtain
\begin{equation} \label{eq-app-psi-e}
\psi_e = (\Omega_{\rm pu} \psi_\uparrow + \Omega_{\rm ca} a \psi_\downarrow)/\Delta_e \,.
\end{equation}
Submitting Eq.(\ref{eq-app-psi-e}) back to Hamiltonian (\ref{eq-app-h-1}) gives
\begin{align}
&\mathcal{H}=(-\Delta_{\rm ca}+\mathcal{E}_{\rm ca}) a^\dag a + \int  d \bfr \,\{ \Gamma_\downarrow \psi_{\downarrow}^\dag(\bfr) \psi_{\downarrow}(\bfr) \notag\\
&+ \mathcal{E}_{\rm pu} \psi_\uparrow^\dag(\bfr)\psi_\uparrow(\bfr)
+[ g(x)a\psi_\uparrow^\dag(\bfr)\psi_\downarrow(\bfr)+H.c.] \} \,, \label{eq-app-h-2}
\end{align}
where we denote the Stark shifts $\mathcal{E}_{\rm pu}=2|\Omega_{\rm pu}|^2/\Delta_e$
and $\mathcal{E}_{\rm ca}=\frac{2}{\Delta_e}\int |\Omega_{\rm ca}(x)|^2\psi_\downarrow^\dag(\bfr)\psi_\downarrow(\bfr)  d \bfr$.
The atom-cavity coupling mode is presented by $g(x)=2\Omega_{\rm pu}^*\Omega_{\rm ca}(x)/\Delta_e\equiv g\cos(k_c x)$
with $g\equiv 2\Omega_{\rm pu}^*\Omega_{\rm ca}/\Delta_e$.

In practice, $\mathcal{E}_{\rm ca}$ is ignorable if $\mathcal{E}_{\rm ca}\ll\Delta_{\rm ca}$.
We can tune $\mathcal{E}_{\rm pu}\approx \Gamma_{\downarrow}+2U_{\rm ac}$,
where $2U_{\rm ac}$ is originated from the first term of $H_{\rm int}^{(0)}$ in Eq.(\ref{eq-h-int-k-each}).
In this way, the energy difference between opposite spins is compensated and thus neglected.
The atom-cavity coupled Hamiltonian (\ref{eq-app-h-2}) reduces to the form: $\mathcal{H}=H_{\rm c}+H_{\rm ac}$ 
(see Eqs.(\ref{eq-h-start-c}) and (\ref{eq-h-start-ac})).

\section{Derivations to Hamiltonian (\ref{eq-h-int-k-each})} \label{sec-app-int}

In the presence of a lattice trap,
we use the tight-binding approximation by expanding $\psi$ in terms of the Wannier basis $W(x)$,
\begin{equation}
\psi_{\sigma}(\bfr) = \sum_j W(\bfr-\bfr_j) \psi_{j\sigma} \,,
\end{equation}
where $\bfr_j$ is the coordinate of the $j$-th site.
Under the Fourier transformation,
$H_{\rm aa}$ in Eq.(\ref{eq-h-eff-1}) is rewritten as
\begin{equation}
H_{\rm aa} = -U_{\rm fr} \sum_{\bfk,\bfk',{\bf p}}
\psi_{\bfk'-{\bf p}\uparrow}^\dag \psi_{\bfk+{\bf p}\downarrow}^\dag\psi_{\bfk\downarrow}\psi_{\bfk'\uparrow}\equiv H_{\rm int}^{(1)}
\end{equation}
with $U_{\rm fr}=g_{\rm fr}\int |W(\bfr)|^4 d \bfr$.
Notice that
\begin{equation} \label{eq-app-gx}
\int  d\bfr\,g(x)\psi_\sigma^\dag(\bfr)\psi_{\bar{\sigma}}(\bfr)=\frac{\eta g}{2}\sum_{\zeta=\pm}\psi_{\bfk+\zeta\bfk_c\sigma}^\dag\psi_{\bfk\bar{\sigma}}
\end{equation}
with $\eta = \int e^{i\bfk_c\cdot\bfr}|W(\bfr)|^2 d\bfr$. By using Eq.(\ref{eq-app-gx}),
Eq.(\ref{eq-h-int-ac}) is rewritten as
\begin{equation} \label{eq-app-h-int-ac}
\widetilde{H}_{\rm ac} = U_{\rm ac}\sum_{\bfk,\bfk'}\sum_{\zeta,\zeta'}\psi_{\bfk+\zeta\bfk_c\uparrow}^\dag\psi_{\bfk\downarrow}
\psi_{\bfk'+\zeta'\bfk_c\downarrow}^\dag\psi_{\bfk'\uparrow}
\end{equation}
with $U_{\rm ac}= \eta^2g^2/(4\Delta_{\rm ca}$).
Due to the anticommutation of $\psi$, we can finally obtain the terms $H_{\rm int}^{(0)}$, $H_{\rm int}^{(2)}$, and $H_{\rm int}^{(3)}$ from Eq.(\ref{eq-app-h-int-ac}).

\bibliographystyle{apsrev4-1}
\bibliography{bib}
\end{document}